# Reduction of optical reflection from InP semiconductor wafers after high-temperature annealing

Oleg G. Semyonov, Arsen V. Subashiev, Alexander Shabalov, Nadia Lifshitz, Zhichao Chen, Takashi Hosoda, and Serge Luryi

*State University of New York at Stony Brook, Department of Electrical and Computer Engineering, Stony Brook, NY 11794*

PACS 78.30.Fs,78.55.Cr,78.60.Lc,74.62.En

We observed and studied strong reduction of optical reflection from the surface of InP wafers after high-temperature annealing. The effect is observed over a wide range of the incident wavelengths, and in the transparency band of the material it is accompanied by increasing transmission. The spectral position of a minimum (almost zero) of the reflection coefficient can be tuned, by varying the temperature and the time of annealing, in the spectral range between 0.5 and 6 eV. The effect is explained by the formation of a uniform oxide layer, whose parameters (thicknesses and average index) are estimated by detailed modeling.

**Keywords**: antireflection, semiconductors, high-temperature annealing, oxide film

## Introduction

Various methods of light reflection reduction from the surface of semiconductors have been intensely studied recently. The antireflection effect is beneficial for many devices such as solar cells, LEDs, and sensors. Routinely, the coatings of a lower refractive index or with a gradient refractive index are used for antireflection.[1] A different approach to antireflection is to create the surface relief with sub-wavelength lateral size of the pattern.[2, 3] Yet another method is based on metallic nanoparticles disposed on the surface.[4] In the latter case, the antireflection effect can be significant even if the thickness of the layer containing the nanoparticles is much smaller than the wavelength of the incident light. This approach, often referred as ‘plasmonic’, has been used, in particular, to render the nearly perfect IR absorbers [5] and other devices with suppressed surface reflection.[6] As a rule, the metallic nano-beads of a particular size and form are positioned periodically on the surface with a specified distance between the beads. As recently shown, the metallic nanoparticles of various sizes on a flat or textured surface can also produce the broad-band antireflection effect.[7] The coatings and structures on the surface are rendered artificially using sputtering and/or lithographic technique.

The reflection reduction can also be caused by surface modification. It is a common experience that surface layers of semiconductor wafers are significantly modified in the process of high-temperature annealing. For InP, the commonly known effect is depletion of phosphorous in the surface layers due to outdiffusion.[8] Relatively less known is the effect of indium diffusion and the formation of metallic nano-clusters or droplets on the surface, observed experimentally after high-temperature annealing of InP wafers in the etching ambient.[9] Oxidation of the surface layers of InP under high-temperature annealing in oxygen ambient has also been studied.[10,11] The reported refractive indices range from 1.4 given in Refs [12,13] and cited in the optical databases [14] for indium oxide sputtered on the surface or deposited electrochemically to the value of 1.9 measured in thermal oxides.[11,15] Either way, the refractive index of oxide layers is lower than that of the wafer material and that can cause suppression of the reflection.

This paper studies the reflection and transmission spectra of n-type InP wafers after their annealing in air at different temperatures. We found a strong reduction of reflection in a broad spectral range. Analysis of the spectra enabled us to estimate the oxide layer thickness and refraction index and make some conclusions about its composition.

## 1. Experiment

Two 350 μm-thick n-InP wafers manufactured by Nikko Metals Inc.[16] doped with sulfur with the carrier concentration of $2\times10^{17}$ cm$^{-3}$, both sides electro-chemically polished, were cut into quarters. One quarter of each wafer was kept unprocessed as the reference sample. Three quarters of one wafer were annealed in an oven at 350, 500 and 600°C for 30 minutes under atmospheric pressure and three quarters of the other wafer were annealed at 600°C for 30, 38, and 45 minutes. Additional measurements were done on heavily-doped ($8\times10^{18}$ cm$^{-3}$ ) samples. The raw reflection $R_{\mathrm{tot}}(E)$ and transmission $T_{\mathrm{tot}}(E)$ spectra of all samples were scanned using the Lambda-950 spectrophotometer (Perkin-Elmer) in the wavelength range from 250 nm to 3000 nm (photon energy $6 > E > 0.5$ eV). From the raw data for $R_{\mathrm{tot}}(E)$ and $T_{\mathrm{tot}}(E)$, the reflection coefficient $R(E)$ and the transmittance $t = \exp[-\alpha(E)d]$, where $d$ is the wafer thickness and α the absorption coefficient, were numerically calculated, accounting for the multiple reflections of the probe light beam from both sides of the samples:[17]

$$R_{tot} = R\left(1+\frac{(1-R)^2 t^2}{1-R^2 t^2}\right), \qquad T_{tot} = \frac{(1-R)^2 t}{1-R^2 t^2} \tag{1}$$

Flipping the wafers upside down results in same raw curves, therefore both surfaces are assumed to have the same reflection coefficient $R(E)$.

The reflection coefficient energy dependence for a reference sample and the samples annealed at 350, 500 and 600° C for 30 min are shown in Fig. 1. Also shown is the reflection spectrum for a heavily doped sample annealed at 900° C, as well as the InP reflection spectrum from Ref. [18]

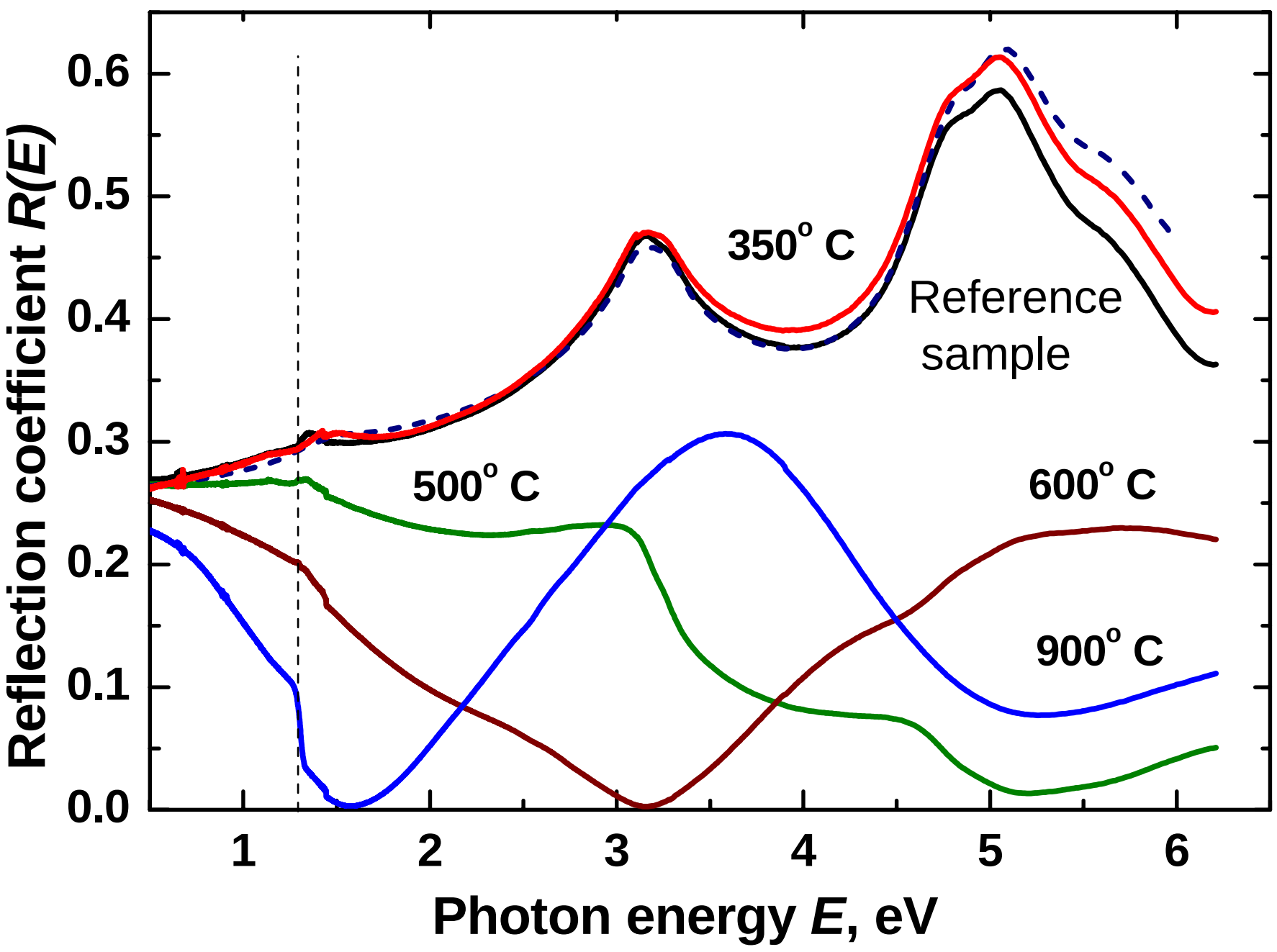

Fig. 1 Reflection coefficient as a function of the photon energy for the reference InP sample and the samples annealed at 350 (red), 500 (green), 600° C (brown) for 30 min and at 900° C (blue) for 5 min. The InP spectrum [18] is also plotted for comparison (dashed curve). The vertical dashed line indicates the position of the absorption edge.

Strong reflection reduction in a broad energy range for annealed InP wafers is observed when the annealing temperature exceeds 350° C. A deep minimum on the reflection graphs is observed for the anneal temperatures 500° C and 600° C (30 min annealing time) with the minimum values of the reflection coefficient $R_{min}$ = 0.013 and 0.002 for the photon energies $E_{min}$ = 5.2 eV and 3.19 eV, correspondingly. For even higher temperatures, the spectral minimum shifts further to the infrared and an additional minimum appears in the ultraviolet (see the 900° C curve corresponding to a much shorter annealing time with faster oxide growth).

Details of the variation of $R_{tot}(E)$ and $R(E)$ in the vicinity of the interband absorption edge, are shown in Fig. 2, together with the corresponding graphs for $T_{tot}(E)$ and $t(E)$.

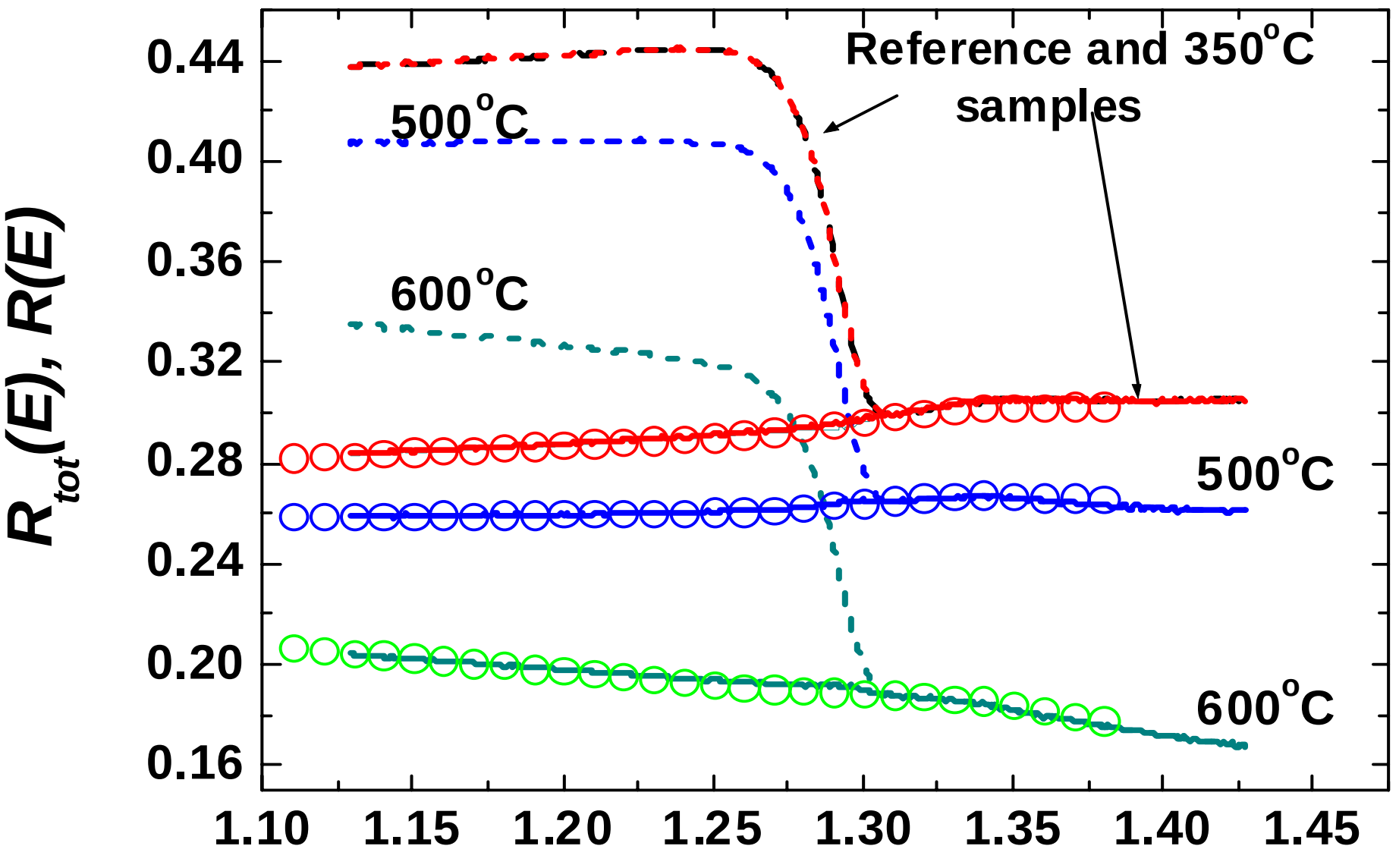

a)

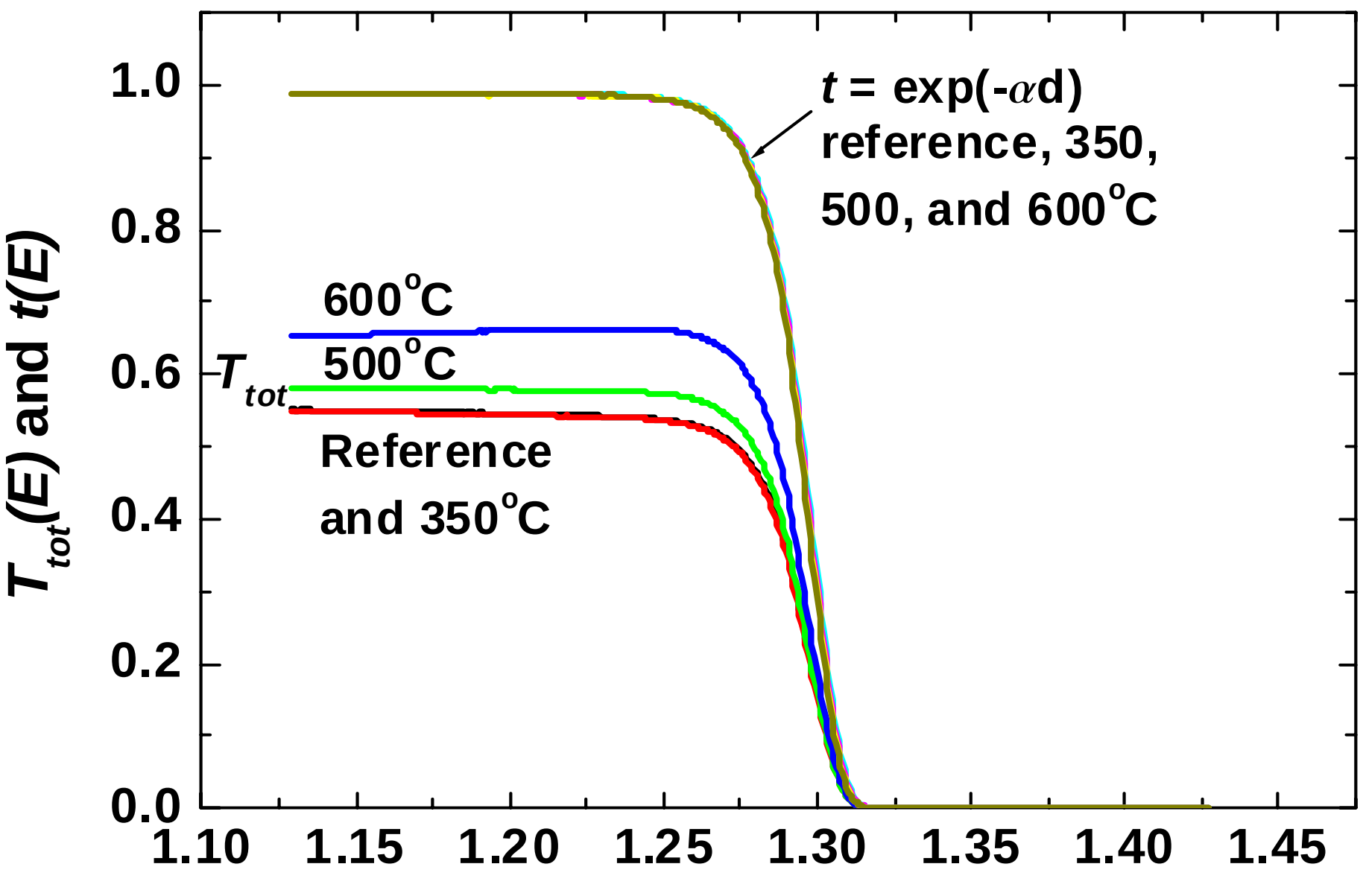

b)

Fig. 2 (a) Raw total reflection $R_{tot}(E)$ (dashed curves) and reflection coefficient $R(E)$ (solid curves) in the spectral band $1.1 < E < 1.5$ eV of the reference sample and the samples annealed at three different temperatures ($N_D = 2\times10^{17}$ cm$^{-3}$, see Fig. 1) for 30 minutes. Also shown are the reflection coefficients $R(E)$ calculated from Eq. 2 (circles, see Discussion). (b) Raw transmission spectra $T_{tot}(E)$ and transmittance $t(E)$ of the samples at the same annealing temperatures.

For 30-min annealing at 500 and 600° C, the reflection is steadily reduced, causing an increase of the total transmission, while the transmittance in the transparence band does not virtually differ from the transmittance of the reference sample. The very high ($t \approx 0.99$) transmittance is owing to the low ($\alpha < 0.3$ cm$^{-1}$) residual free carrier absorption in low-doped InP wafers.

Figure 3 shows the spectral dependence of the reflection coefficients $R(E)$ of InP samples annealed at T = 600° C for 30 min, 38 min, and 45 min. With the increasing anneal time the reflection minimum progressively shifts from ultraviolet to infrared, from 3.19 eV ($R_{min} = 0.002$) for 30-min annealing to 1.14 eV ($R_{min} = 0.027$) for 45-min annealing. The latter value lies in the sample's transparency region. For the annealing times longer than 30 min, one sees in the ultraviolet an additional minimum, which resembles the next interference minimum for an oxide layer on the surface. This second minimum is, however, spectrally shifted due to the dispersion in the refractive indices and the absorption coefficients of both the wafer material and the layer itself.

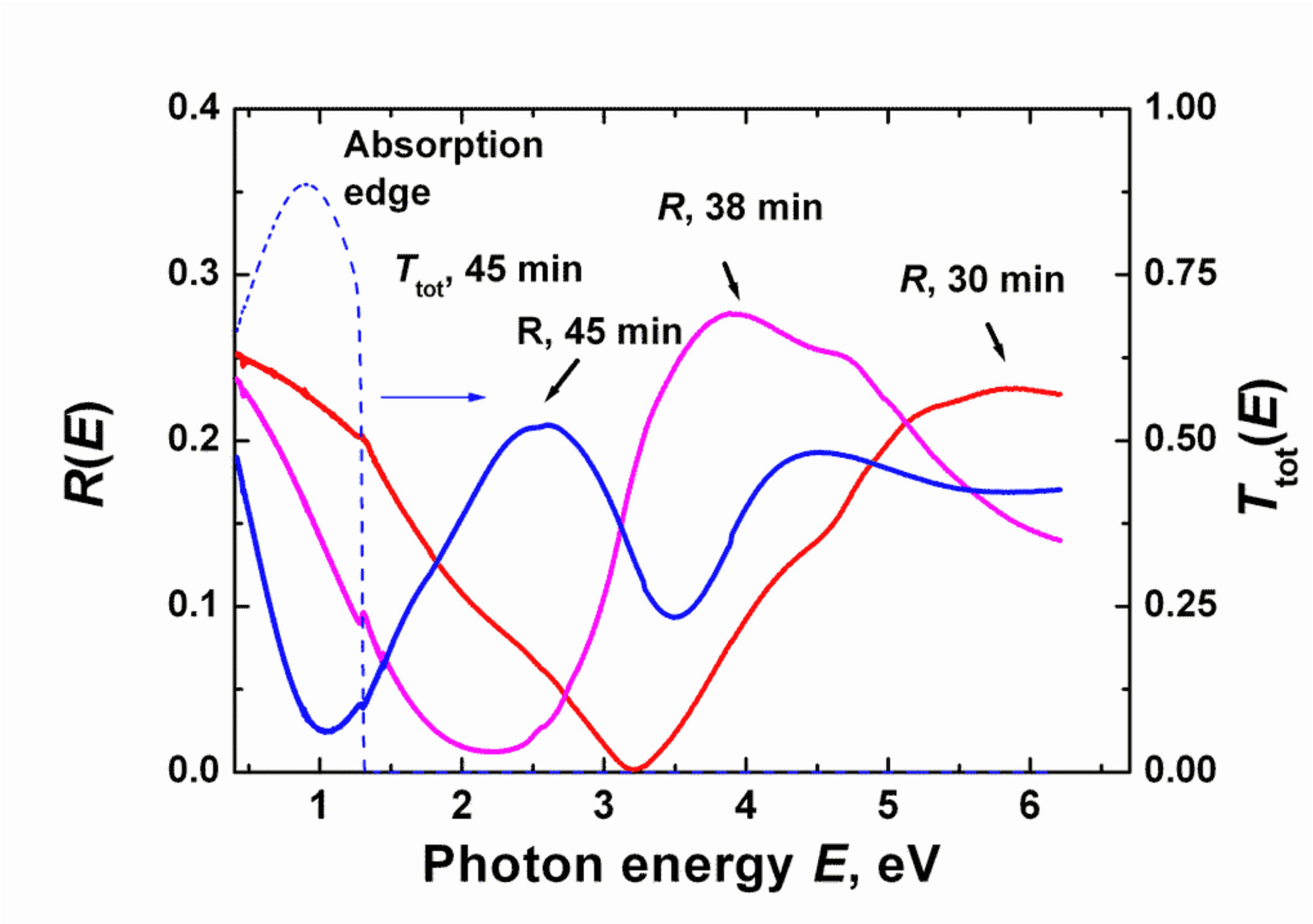

Fig. 3 Reflection coefficient $R$ as a function of the photon energy for InP samples annealed at 600° C during 30 min (red), 38 min (magenta), and 45 min (blue). Also shown (dashed curve) is the raw transmission curve $T_{tot}$ for the annealing time of 45 minutes showing a dramatic increase of transmission of the sample in the transparency band due to the decrease of reflection.

Also shown in Fig. 3 is the transmission $T_{tot}(E)$ of the wafer in the transparency region after 45-min anneal. It demonstrates the increase of transmission that accompanies the reflection reduction. The calculated transmittance of this sample reaches $t \approx 98.8\%$ in the transparency band of InP, with some scattered light of the laser light (980 nm) seen on the surface through a night-vision IR viewer. Visually, the surface color of the annealed wafers changes in comparison with the untreated samples: the surface of a virgin wafer looks shiny dark-gray while the surface of the sample annealed for 30 minutes at 600° C becomes brownish and the surface of the sample annealed for 45 minutes at 600° C attains a slightly bluish hue.

We have also measured the transmission of a 980-nm laser beam (in the transparency region of InP) through the sample annealed for 45-min at 600°C as a function of the angle of incidence separately for *s* and *p*-polarization (Fig. 4).

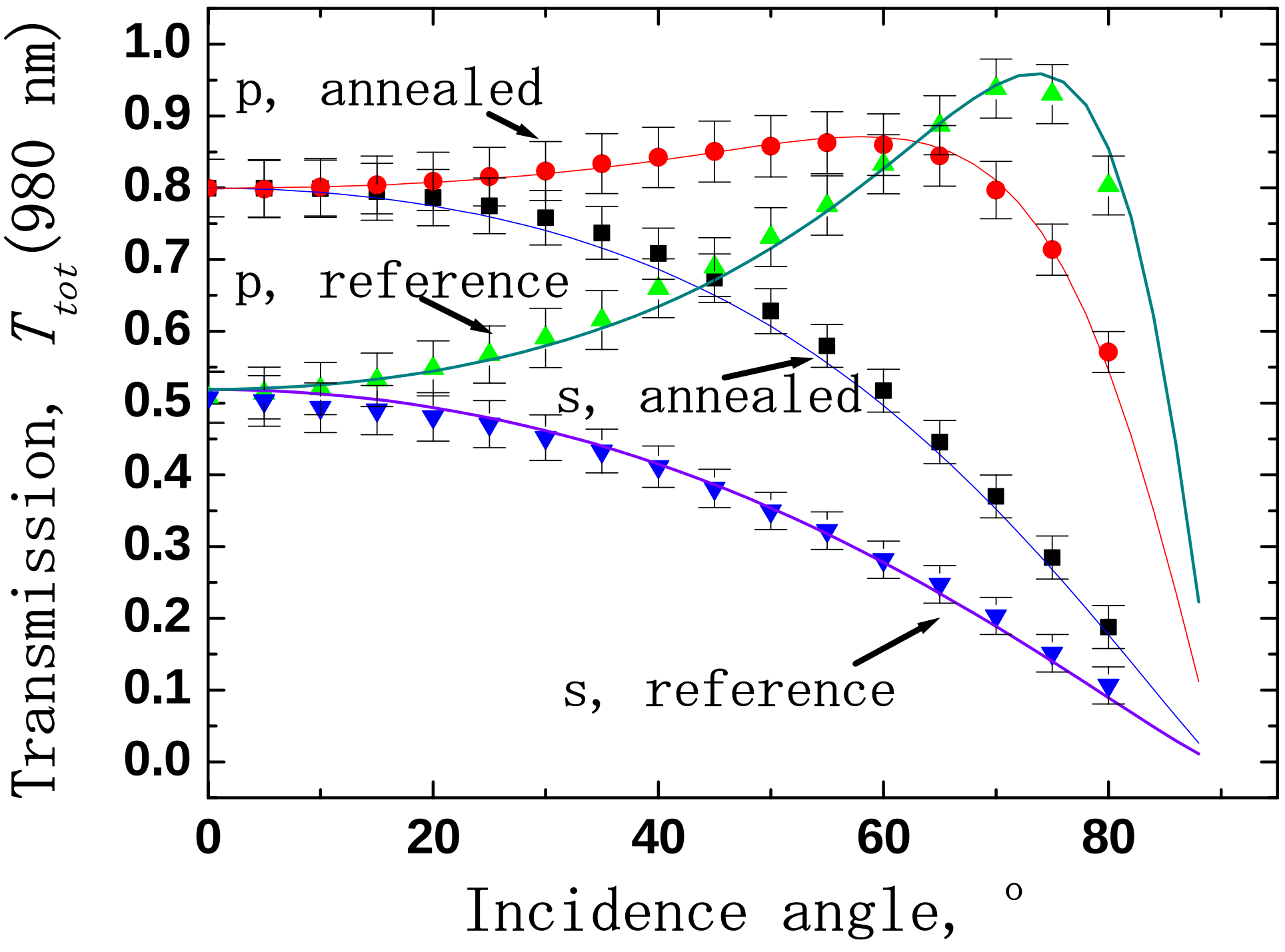

Fig. 4 (Color online). Transmission $T_{tot}$ of the reference sample and the sample annealed at T = 600° C for 45 minutes as a function of the incidence angle for s and p-polarizations of the incident laser beam of 980 nm ($E$ = 1.267 eV) (dots). Solid lines show the calculation results (see the text).

For the reference sample, the transmission of *p*-polarized light grows with the angle of incidence toward a maximum near the Brewster angle at φ ≈ 73° which corresponds to the refractive index of InP $n$ = 3.36, while the transmission of *s*-polarized light uniformly declines with the angle. For the sample annealed at 600°C for 45 minutes, the $T_{tot}(\varphi)$ of *p*-polarized light increases toward a pseudo-Brewster angle that is shifted to φ ≈ 55°. Also shown are theoretical curves calculated with Fresnel's equations for a homogeneous oxide layer with the refractive index of 1.95 (see Discussion).

It is well known that high-temperature annealing can alter the InP surface [8-11]. Surface scans of our samples with the Veeco Dimension atomic force microscope (AFM) shows noticeable modification of the surface morphology after annealing at temperatures above 350° C (Fig. 5). The reference and 350°C-annealed samples show relatively rare elevations on the otherwise flat surface apparently related to the dust granules on the surface. At higher temperatures, a uniform thicket of spikes is formed over the whole surface. Actually, the surface features resembling spikes in Fig. 5a are relatively gentle hills with their lateral size ~ 50- 100 nm (Fig. 5c) and with the average height increasing with the annealing temperature, from 5-10 nm for 500°C to about 10-17 nm for 600°C (30 min annealing) as seen from the histograms in Fig. 5b. The average lateral size of the 'hills' seems to grow marginally with the annealing time.

Scanning electron microscope (SEM) images of the wafer surfaces annealed at 600° C for different anneal times are shown in Fig. 6. The images of the surfaces and the cleaved sides of the wafers were obtained with an electron beam of 15 keV and 7.5 keV, respectively. The darker areas correspond to the relatively lower effective atomic weight (presumably oxide clusters), while the irregular bright network corresponds to material of higher atomic weight, most likely the substrate surface. The SEM map of the surface for the sample annealed for 30 minutes at T = 600° C revealed the average lateral size of the darker irregularities ~ 50 – 100 nm. The increase of the lateral size of these spots from 50 – 100 nm for 30-min annealing to 0.5 – 1 μm for 45-min annealing can explain appearance of scattered laser light (980 nm) from the surface of the sample annealed for 45 minutes at 600° C as viewed through the night-vision device while no scattered light (only specular reflection) is visible from the sample annealed for 30 minutes at 600°C and from the samples annealed at lower temperatures.

The SEM images of the cleaved edges of the annealed wafers, shown in the bottom parts of the images in Fig. 6, also reveal the darker (presumably, oxide) clusters in the surface layer, with their lateral sizes corresponding to the dark spots of the surface images and with their depth growing with the annealing time. No such patterns are observed in the reference samples. X-ray microanalysis of the cleaved edge cross-section beneath the dark-spot layer showed no sign of oxygen while in the layer itself the oxygen content is high (Fig. 7).

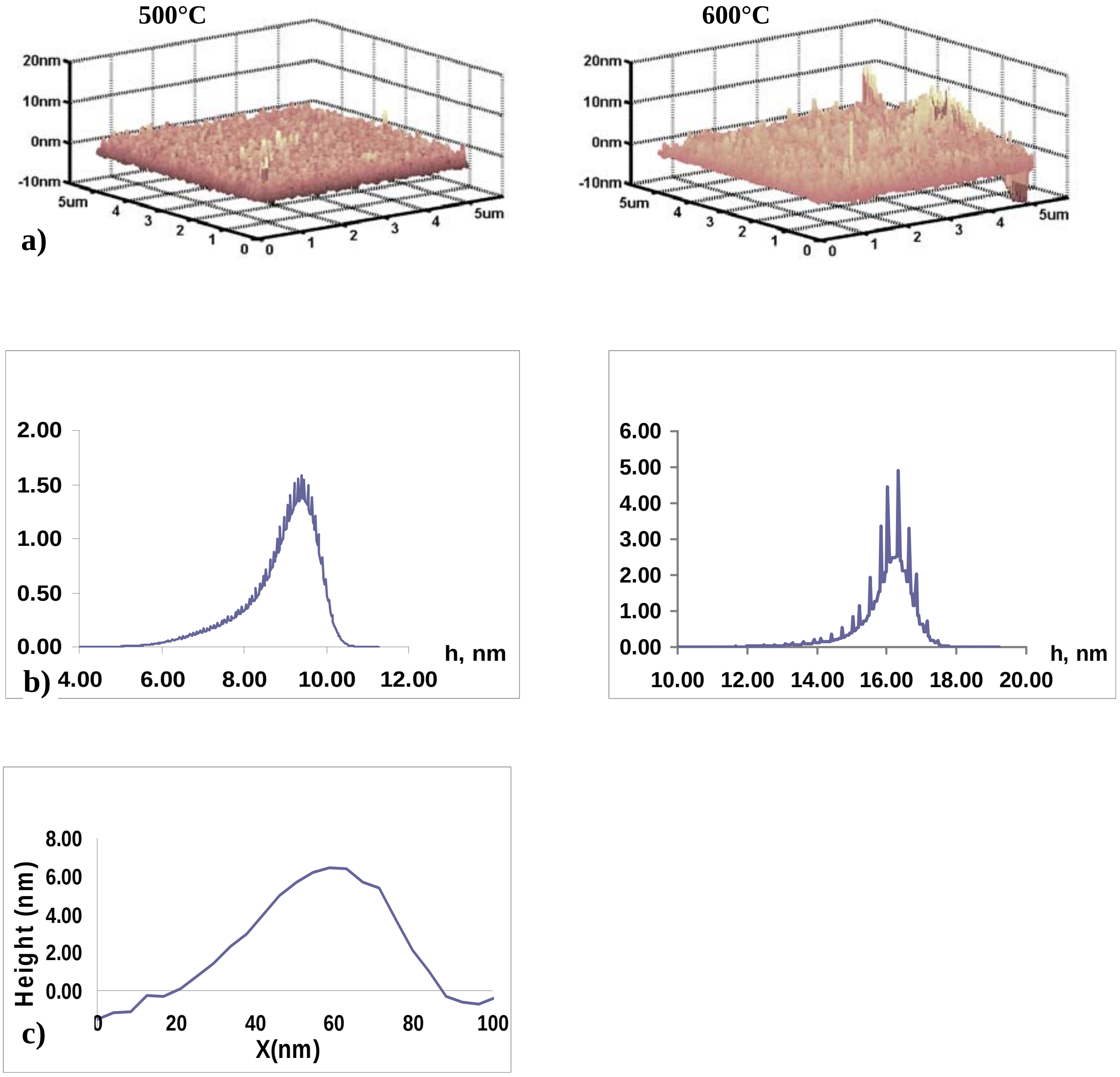

Fig. 5 AFM scans of portion of InP wafers annealed at 500° C (left) and 600° C (right) (a), the corresponding histograms of the surface relief distribution in the height h of the spikes (hills) (b), and the scans of a single representative ‘hill’ of the relief for 500° C annealing temperature (c).

Oxidation seems to begin in small local patches of InP near the surface and then the formed oxide clusters grow just beneath the surface with temperature and annealing time (compare the side SEM images for 30 min and 38 min anneals). Remarkably, the surface relief of the wafers obtained with 2.5-keV electrons does not reveal this patterning and the typical sizes of the surface relief are virtually close to the patterns obtained with AFM scans. It means that the oxidation penetrates into InP and results in a growing depth of the oxide clusters while the surface relief is subjected to relatively minor changes. This can be also seen from the images of the cleaved edges of the samples in Fig. 6.

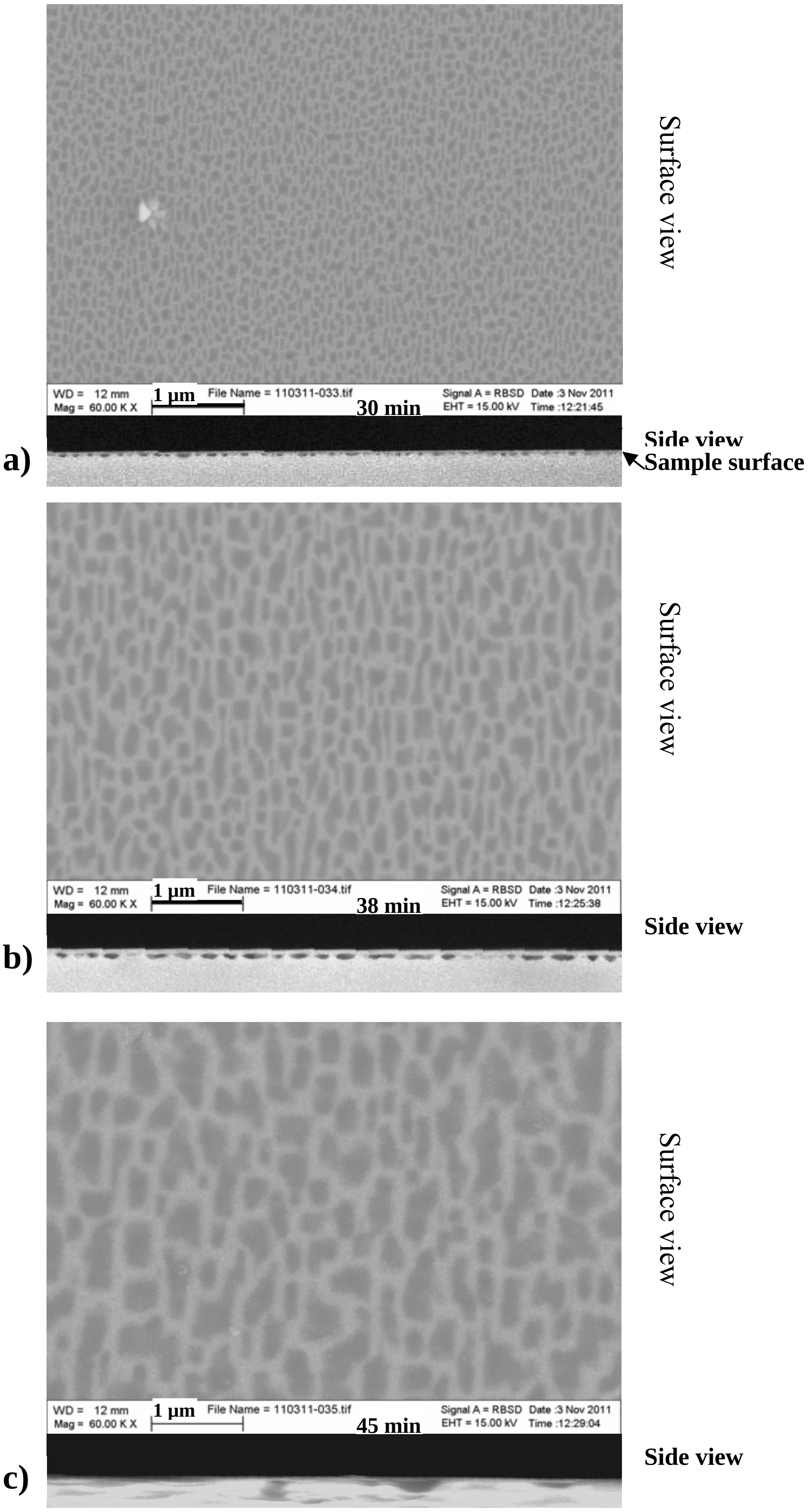

Fig. 6 SEM images of the samples annealed at 600° C, for 30 (a), 38(b), and 45(c) minutes, respectively. The darker areas correspond to the patterns of lower density in terms of the effective atomic weight (presumably, oxide clusters). Note that their lateral size grows with the annealing time from 50-100 nm (a) to 0.5-1 µm (c). The corresponding SEM images of the cleaved sides of the samples are shown at the bottom section of each micrograph.

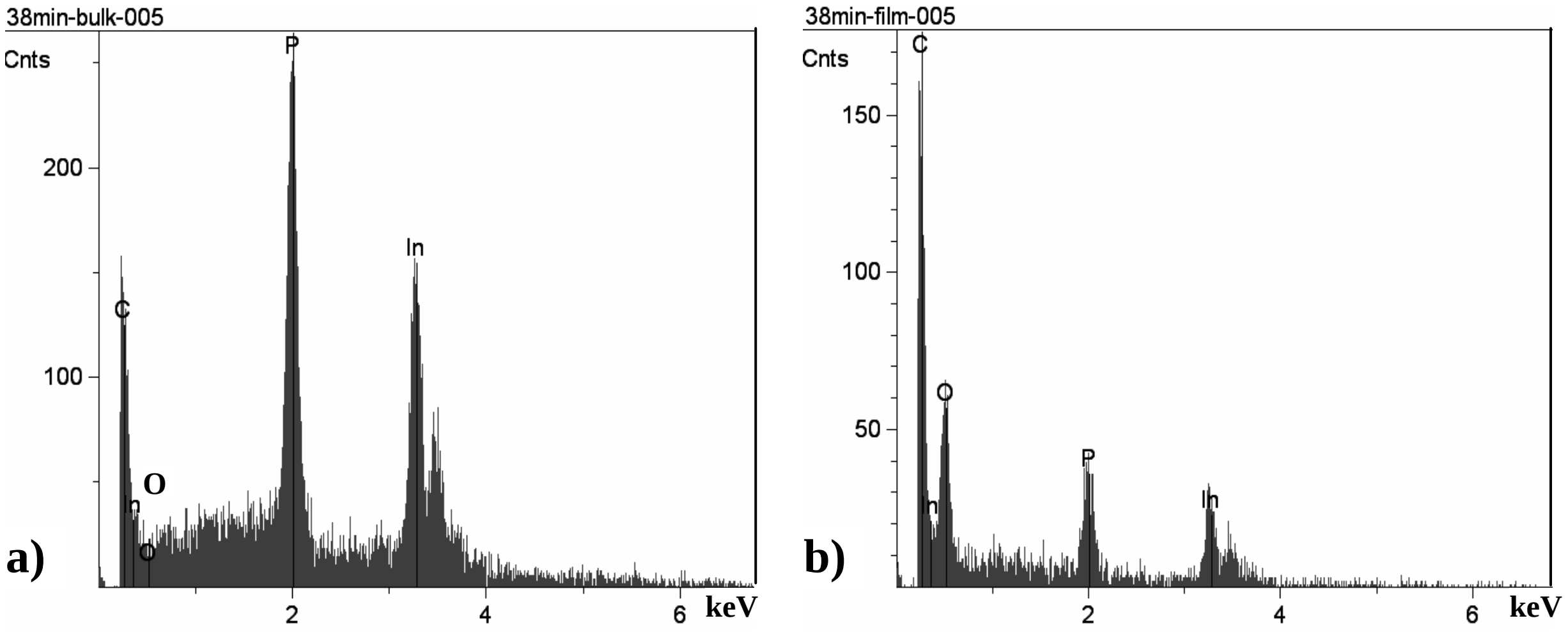

Fig. 7 Characteristic X-ray spectra from the cleaved edge of the sample annealed at 600° C for 38 minutes: a) the uniform bulk area well beneath the oxide layer; b) the oxide layer with the darker patterns (oxide clustering) near the surface of the sample (see the bottom part of Fig. 6b). The counting time for the oxide spectrum (b) significantly exceeds the counting time for InP spectrum because of the smaller radiation collection area. The peak at the lowest energy marked by C originates from the microscope background (organic contamination).

The described antireflection effect associated with surface oxidation is not exclusive to InP but is also observed with other semiconductors. In particular, we have observed the anneal-induced antireflection effects in GaAs and GaP wafers (Fig. 8). Similar to our procedure with InP, we cut these wafers into several samples and one quarter of each wafer was kept unprocessed (reference sample). For GaP samples the substantial reflection reduction effect appears at the annealing temperature of 750° C (no changes at 600° C and below) and the minimum of reflection also shifts to lower energies with the increasing annealing time. For GaAs, the antireflection is observed at 600° C after 30 minutes anneal (Fig. 8b) but in contrast with InP and GaP, the reflection minimum shows no redshift with the excess annealing times above 30 min.

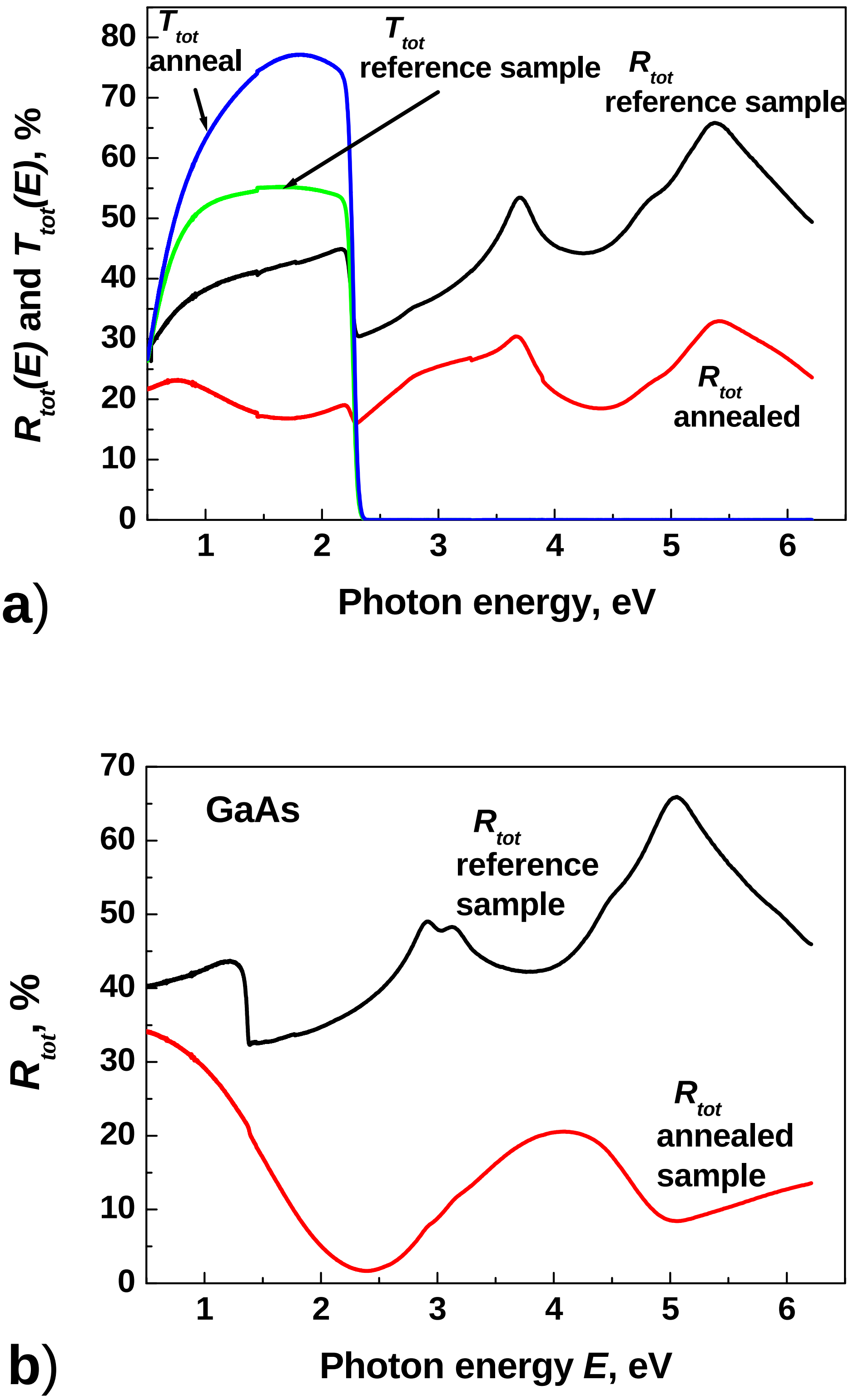

Fig. 8 Reflection reduction in GaP wafer (a) and the antireflection minimum in GaAs wafer (b) after high-temperature annealing. The much stronger effect in GaAs correlates with the higher index ($n \approx 1.9$) of its oxide, compared to $n \approx 1.6$ for GaP oxide.[14]

## 2. Discussion

The reflection and transmission spectra were analyzed with different assumptions about the surface oxide layer structure. The notable feature of the experimental spectra is the almost complete suppression of reflection at a particular photon energy and the shift of the spectral position of the reflection minimum to lower photon energies with increasing temperatures and annealing times. We have found that a homogeneous oxide-layer model can reproduce these features. In order to get nearly total antireflection, two conditions must be met: (i) the phase shifts for light waves reflected by different interfaces must be matched; and (ii) the index of refraction $n$ of the layer must be close to the square root $(n_s)^{1/2}$ of the substrate refractive index. To calculate the reflection spectra for a homogeneous oxide layer, we used the conventional Fresnel approach with the energy dependent complex refractive index $n + i\kappa$ of InP. [18] Our calculations showed that the antireflection condition (ii) may be rather relaxed: an oxide layer with index in the range 1.75 to 2 and a thickness above 20 nm will lead to a reflection minimum with $R_{min} \leq 0.2\%$. The strong variation of $n$ and κ of InP with the photon energy narrows the range of suitable refractive indices for the oxide. Thus, the database [14] value ($n \approx 1.4$) for the index of the oxide layer on InP is too low for the observed nearly total antireflection.

If, based on the SEM images, the modified layer were to be interpreted as oxide clusters alternated with InP inclusions, one could estimate the layer dielectric function as an average of the dielectric functions of the oxide and InP, $\langle\varepsilon\rangle = \varepsilon_{InP}\, x + \varepsilon_{ox}\,(1-x)$, where $x$ is the volume content of InP in the layer. This would produce an enhanced oxide index observed experimentally. However, the reflection spectra calculated in this way are at variance with the experiment: the average $\langle\varepsilon\rangle$ would retain some of the characteristic features of $\varepsilon_{InP}$ and produce a markedly different spectral behavior of the reflection coefficient.

Alternatively, we could imagine metallic inclusions corresponding to In embedded in $In_2O_3$ oxide. To calculate the dielectric function in this case, it is appropriate to use the Maxwell Garnett approach, see e.g. Ref. [20]. This would also produce an enhanced refractive index. However, thus calculated reflection spectra would also be at variance with the experiment: the contribution of metallic particles to the total dielectric function would contribute a characteristic plasmonic dispersion, which is not observed.

The model of a homogeneous oxide layer yields a much better agreement with the experiment. The results of our calculations for the samples annealed at 600° C are shown in Fig. 9. For 30 min and 38 min annealing times, the closest fit was obtained with both the real part $n$ and the imaginary part $\kappa$ of the refractive index dependent on the photon energy $E$. We find $n(E) = 1.95 + 1.4\times10^{-3}E^{2} + 4\times10^{-5}E^{4}$ and $\kappa(E) = 2.8\times10^{-2}(E-1.5)^2$, where $E$ is in eV. Our values of $n$ are larger than 1.6 quoted by Studna and Gualtieri [12] but close to 1.9 cited by Nelson at al. [11]

Our spectra for $n(E)$ are reasonably close to those measured by Robach et al.[15] where $n(E)$ ranged from 1.85 to 2. To fit the reflection curve for the sample with the annealing time of 45 min, the index $\kappa$ is taken as $\kappa = 3.6\times10^{-2}\,(E–0.5)^2$, which corresponds to higher oxide layer absorption in this sample. In the reflection spectra of Fig. 9 the higher $\kappa$ is manifested by a shallower interference pattern above 3 eV. Quadratic growth of absorption is typical for amorphous layers. [19] The increasing thickness of the oxide layer is, apparently, accompanied by some change in its composition resulting in enhanced absorption. The calculated thickness $d$ of the oxide layer varies from 44 nm for the sample annealed for 30 min to 120 nm for the sample annealed for 45 min. The layer is thick enough to provide a strong antireflection effect when its optical thickness is close to λ/4 (some deviations from the condition $nd = \lambda/4$ may be linked with absorption in the InP substrate which causes an additional phase shift). The increase of the oxide layer refractive index compared to that for $In_2O_3$ index may be due to inclusions of the dielectric compounds $InPO_4$ and $In(PO_3)_3$, manifested in the X-ray spectra [11, 15].

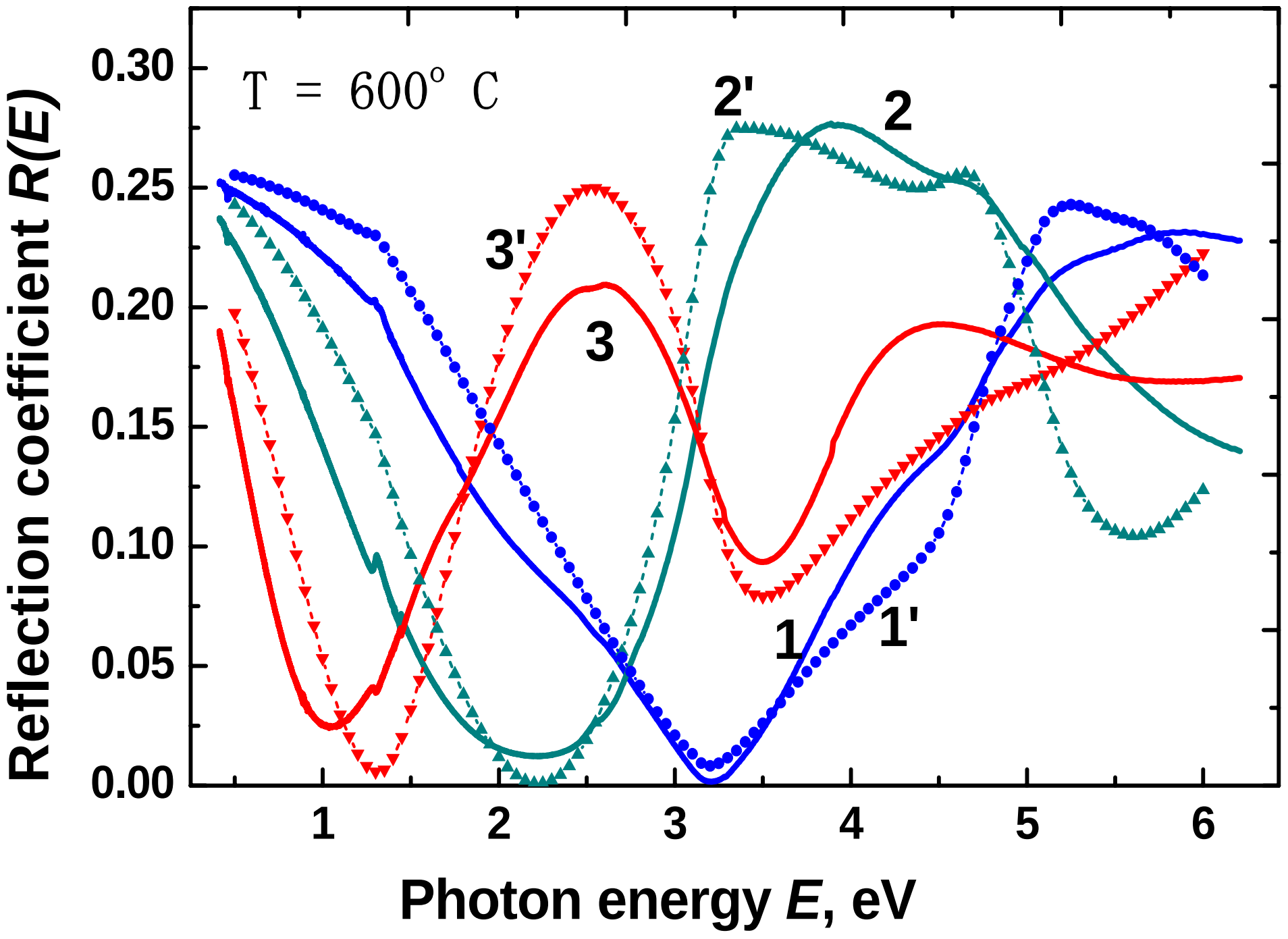

Fig. 9 (Color online). Experimental data (solid curves 1, 2, 3) for the annealing times of 30 (blue), 38 (green), and 45 (red) minutes and their fitting (dotted curves 1′, 2′, and 3′) by the Fresnel formula for the homogeneous oxide layers of $d$ = 44 (blue), 68 (green), and 122 (red) nm, respectively.

The lower annealing temperatures result in reduced oxidation rates and hence thinner oxide layers, formed during a fixed annealing time. The experimental reflection spectra, for two anneal temperatures of 500° C and 600° C and the same annealing time of 30 min, are shown in Fig. 10 and compared with several models of the oxide layer. For T = 500°C, the reflection spectrum has a characteristic structure inherited from the refractive index variation of InP with a deep antireflection minimum near $E$ = 5.2 eV. The thickness of the oxide layer estimated from the condition $d \approx \lambda/4n$ is 28 nm for $n = 2$. At T = 600°C, the reflection minimum shifts to $E$ = 3.2 eV, which corresponds to thickness $d \approx 48$ nm. The reflection spectra calculated in the model of a homogeneous oxide layer are shown in Fig. 10 by the dashed lines (1′ and 2′). The best fit gives slightly smaller thicknesses of the oxide layer than those estimated above. The difference is due to an additional phase gain in the reflection from an interface between the oxide and the absorbing wafer. The model of a homogeneous oxide layer also gives an excellent account to our results for 500°C annealing temperature. The best fit is obtained with a slightly lower absorption index of the oxide layer, viz. $\kappa = 1.7\times10^{-2}(E-1.5)^2$ and a correspondingly slightly smaller dispersion of the real index at high energies, $n = 1.95 + 1.4\times10^{-3}E^2 + 1.5\times10^{-5}E^4$.

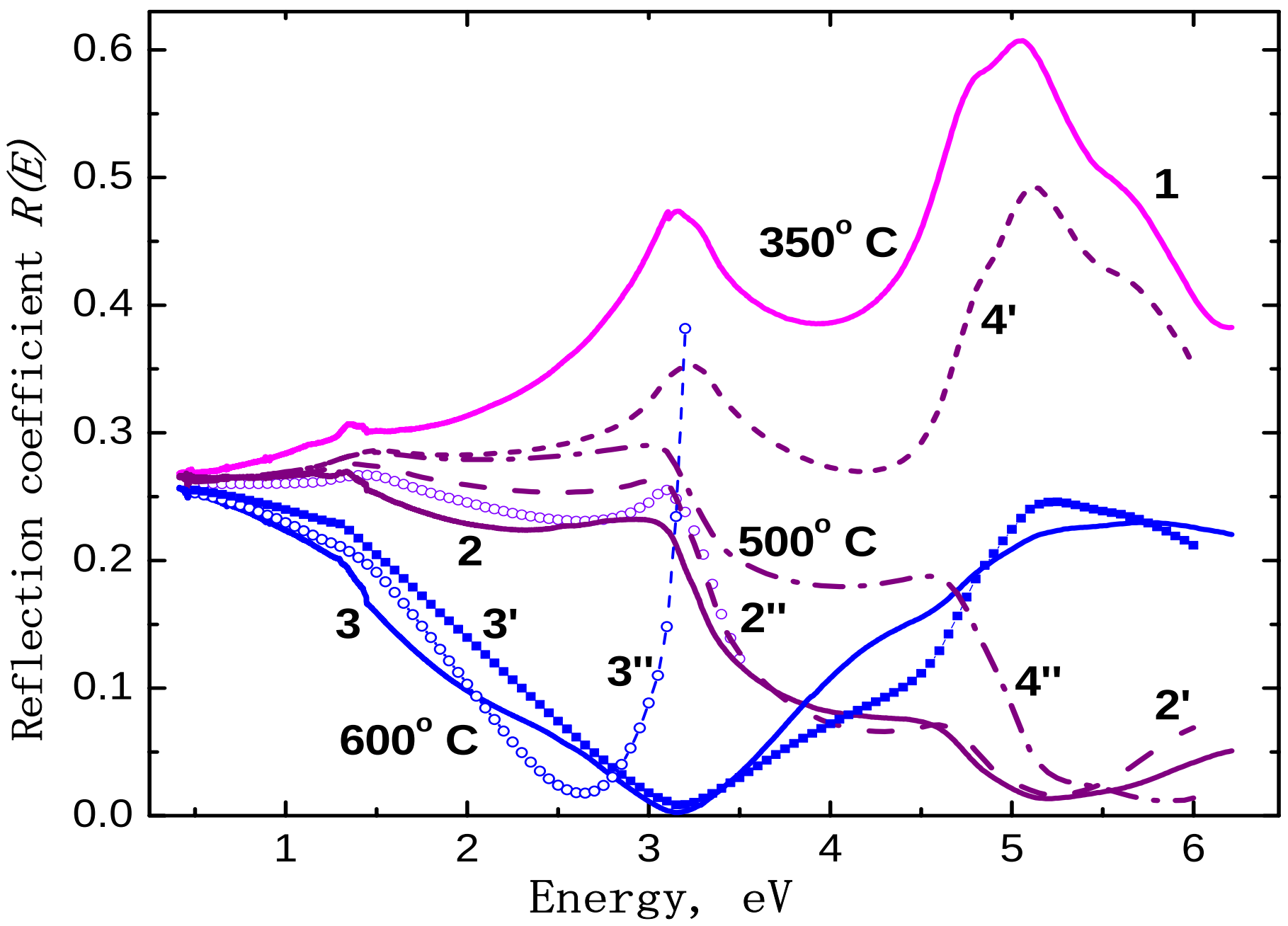

Fig. 10 Thin solid curves 1, 2 and 3 show the experimental reflection coefficient $R(E)$ for the annealing temperatures 350° C (red), 500° C (brown) and 600°C (blue), respectively. The dashed curve 2′ (brown) and the curve 3′ with square dots show the calculated results for a homogeneous oxide layer of thickness 22 nm (2′) and 44 nm (3′); the curves with open dots 2″ (brown) and 3″ (blue) are the calculated $R(E)$ characterized by a certain surface invariant $I_\tau$ according to Bedeaux at al. [20] for the same effective thicknesses of the layer [22 nm (2″) and 44 nm (3″)]; the short-dash curve 4′ (brown) and the dash-dot curve 4″ (brown) are the spectra calculated from the Epstein model [21] for the textured InP

The homogeneous layer model also gives a good fit to the angular dependence of the total transmission in the transparency region, as shown in Fig. 4. Note that the angular dependence of *p*-polarized wave is sensitive not only to the effective optical thickness *nd*, but to the index *n* and the thickness *d* of the interlayer individually. An excellent agreement was obtained with the experimental results, as shown in Fig. 4, yielding $d$ = 100 nm and $n$ = 1.95. The refractive index is identical to that estimated earlier from the experimental reflection spectrum in the wide energy range. On the other hand, the thickness *d* is somewhat smaller than 120 nm estimated earlier.

The homogeneous layer model yields rather large thicknesses of the oxide layers (over 100 nm for 45-min annealing time), which is hard to confirm by direct observation. For comparison, we have calculated the reflection coefficient *R(E)* in the Bedeaux model,[20] which is well suited to describe the antireflection effect for thin ($d \ll \lambda$) surface layers on a substrate with a higher refractive index. The model is based on the description of the surface layer through its polarizability expanded in multipole moments. It results in a modified Fresnel formula for the reflection coefficient. In the transparency region of both the interlayer and the substrate and for the normal incidence it is of the form:

$$R = \frac{(1-n)^2 - 2k^2 I_\tau (n^2-1)}{(1+n)^2 - 2k^2 I_\tau (n^2-1)}, \qquad (2)$$

where $k = 2\pi/\lambda$. Compared to the conventional Fresnel equation, an additional term is included in both the numerator and the denominator of Eq. (2). For the normal incidence this additional term is proportional to the invariant parameter of the surface layer $I_\tau$ (invariant with respect to the choice of the wafer surface position) which can be expressed in terms of the quadrupole moment of the oxide layer.[20] For a non-absorbing layer, the invariant is real and can be interpreted as a squared effective thickness of the intermediate layer $I_\tau = d_{\mathrm{eff}}^2$. For the surface structure with a large quadrupole moment, the effective optical thickness $d_{\mathrm{eff}}$ of the layer can be much larger than its physical thickness *d*, though these two quantities are proportional to one another. Note that in the transparency region, the correction terms are of the order $(d/\lambda)^2$ and tend to zero with vanishing *d*. In the region of the substrate absorption, the correction terms differ from those in the transparency band: they are larger and proportional to $d/\lambda$.

Using Eq. (2) with the invariant $I_\tau$ as an adjustable parameter, we calculated the reflection coefficient *R*(*E*) in the transparency region for normal incidence. The results shown in Fig. 2 by open circles are in a good agreement with the experimental data for all annealing temperatures. The calculated effective thickness $d_{\mathrm{eff}}$ (T) = $I_\tau^{1/2}$ grows with the annealing temperature and correlates with the results of the homogeneous-layer model. The reflection spectra calculated in the Bedeaux model, [2020] are also plotted in Fig. 10 over a broader range of photon energies (curves 2″ and 3″). One can see that, with a proper choice of the surface invariant, the low-energy reflection coefficient closely fits the experimental data. Remarkably, the Bedeaux model allows one to achieve an almost total antireflection by a relatively thin physical layer endowed with a large quadrupole moment. In this model, the antireflection does not result from interference. Over a broader energy range the Bedeaux model predicts only one deep antireflection minimum, followed by a steep increase of the reflection coefficient. In contrast, the homogeneous layer model produces the second interference minimum in agreement with the experimentally observed $R(E)$. So, the assumption of a thin oxide with an enhanced quadrupole moment, which underlies the Bedeaux model, is ruled out by the observed interference phenomena.

As seen from Figs. 5 and 6, both AFM and SEM scans indicate the presence of roughness on the surface and formation of a structured oxide layer consisting of patches of different effective atomic weight and hence different chemical composition. Therefore, some spatial profile of the refractive index can be expected in the normal direction. To check the impact of the spatial variation of the oxide refractive index, we performed the fit of the experimental reflection curves using the Epstein model [21] with a gradual variation of the refractive index across its profile from $n = 1$ to the value corresponding to the InP substrate. The numerical calculations [22] show that the exact dependence of the dielectric function on the depth (e.g. linear, cubic, tanh-like, or even discontinuous at the layer surfaces) is not important for thin layers so that the only fitting parameter in the model is the layer thickness. The advantage of the Epstein model is that an analytical expression for the reflection coefficient is available. It applies to the case when the material is strongly absorbing and includes the case of a textured wafer. The results of our calculations in this model are also presented in Fig. 10 (curve 4″). For the complex refractive index of textured InP, the agreement with experiment is quite poor. If the absorption is neglected and thus only the real part of index $n$ is used to fit the data (imitating a non-absorbing transition layer with the refractive index varying from $n = 1$ to the index of InP), then the reflection curve for the same oxide layer thickness (25 nm) reproduces the experimental reflection spectrum much better (curve 4′). Still the agreement is not as good as for the homogenous-layer model. Our conclusion is that the observed textured surface is not relevant to the antireflection effect.

From the obtained $n(E)$ and $\kappa(E)$, one can find the energy dependence of the complex dielectric function ε of the surface oxide layer. Its real and imaginary parts, Re ε and Im ε, are shown in Fig. 11 for the annealing temperature T = 600 C$^{o}$. Also shown in Fig. 11 are the data for the thermal oxide layer from Robach et al.[15] Both our results and those of Robach et al. are obtained by solving an inversion problem, which has a limited accuracy. While the absolute values for the real part of the dielectric function are reasonably close, the data of Robach et al [15] show significantly larger variations in Re ε together with a steep decrease of Re ε in the high-energy region, where Im ε continues to grow. This is hard to reconcile with the Kramers-Kronig relation, unless one postulates existence of a peak in absorption just outside the energy range of our experiment. Within our range this would imply a strong increase of absorption with energy, much stronger than the observed quadratic increase. Our data for Im ε are close to those of Robach et al. [15] but our quadratic energy dependence is 'theoretically motivated' and better suited for different annealing conditions.

It may look surprising that a model of a homogeneous oxide layer works so well in a structure with the oxide layers that are manifestly nonuniform in terms of the effective atomic weight, cf. Fig 6. However, the geometry of reflection excitation by a plane wave and observation of a reflected plane wave imposes a self-averaging effect over a large area on the layer surface, so that an average index of the surface layer is measured. The main manifestation of the surface inhomogeneity is the scattering losses. Using an infrared viewer, we indeed observed some scattered light in the 45-min annealed sample irradiated by 980-nm laser beam, though the intensity of the scattered light was rather small. Scattering effects may provide an additional loss mechanism accounting for the slight discrepancy between experimental and calculated spectra.

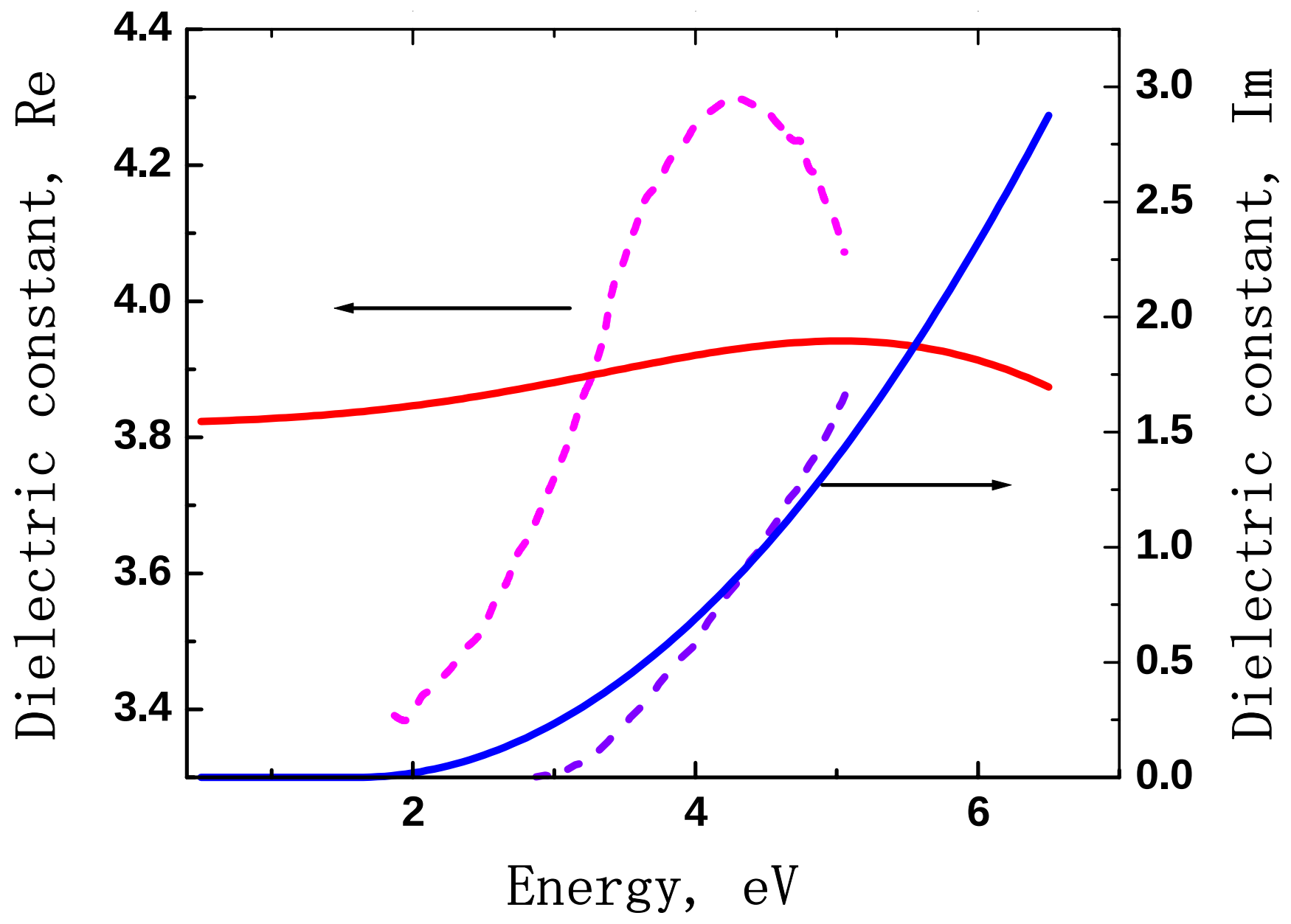

Fig. 11 Dependence of real (red) and imaginary (blue) parts of dielectric function of the oxide layer evaluated from fitting the interference pattern of the reflection spectra at 600 C. Dashed curves show Re ε (magenta) and Im ε (violet) according to Robach et al. [15]

## Conclusion

Antireflection effect of InP samples after their high-temperature annealing in air has been observed. In all cases, the reflection spectra show appearance of a deep minimum on the reflection spectrum. The position of the antireflection minimum is shifted from UV to IR region with the increase of anneal temperature (above 500 C) and/or time (above 30 min). The AFM and SEM microscopy showed that roughness and patterning of the surface layer grow with the annealing temperature and the time of annealing, accompanied by the overall increase of the effective thickness of the layer with time. The thickness and optical homogeneity of the surface layer was estimated by fitting experimental spectra of the reflection coefficient with calculations based on different models of the oxide layer. All models give similar estimates for the oxide layer thickness $d$ that grows with the annealing temperature and annealing time. The closest fit to the experimental spectra in a broad energy range from 0.5 to 6 eV can be obtained with the model of a homogeneous oxide layer having an almost constant real part Re $\varepsilon \approx 3.91$ and an imaginary part Im $\varepsilon$ that grows with the energy as is typical for amorphous layers. The corresponding refractive index is noticeably higher than that of $In_2O_3$ and indicates a more complex composition of the surface layer, with the presence of phosphides.

Our results suggest that studying the reflection in a wide range of photon energies enables determination of the oxide index with higher accuracy – helped by the fact that the dispersion of the refractive index is very different for the semiconductor (high dispersion) and the oxide (low dispersion). We found that the index of thermal oxide on InP is considerably higher than that reported previously and cited in most databases. The new value is in agreement with some earlier publications. The high oxide index gives rise to a deep anti-reflection minimum. Position of the deep minimum can be tuned in a wide range by varying the effective thickness of the oxide layer controlled by the anneal time and/or temperature.

We note that the substantial reduction of reflection is also observed on the surfaces of other $A_3B_5$ semiconductors after high-temperature annealing. The approach used here can be applied to obtain more detailed information about the structure of the modified surface layers.

This work was supported by the Department of Homeland Security through its Academic Research Initiative, by the Defense Threat Reduction Agency through its Basic Research program, and by the NY State Center for Advanced Sensor Technology at Stony Brook.

[13] W.F. Wu and B.S. Chiou, "Effect of annealing on electrical and optical properties of RF magnetron sputtered indium tin oxide films", Applied Surface Science 68, 497-504 (1993).

[14] FreeSnell: Refractive Index Spectra nk-data, http://people.csail.mit.edu/jaffer/FreeSnell/nk.html;

New Semiconductor Materials. Characteristics and Properties, n, k database , http://www.ioffe.ru/SVA/NSM/nk/index.html

[15] Y. Robach, A. Gagnair, J. Joseph, E. Bergignat, G. Hollinger, "Optical properties of native oxides on InP, Thin Solid Films", **162**, 81 (1988).

[16] We used ACROTEC wafers from NIKKO Metals.

[17] Equations (1) are not quite accurate since they do not include the corrections for the phase shift of the reflected wave in the region of the high wafer absorption. However, for our thick samples nontransparent in the absorption region these corrections are negligible.

[18] D.E. Aspnes, A.A. Studina, "Dielectric functions and optical parameters of Si, Ge, GaP, GaAs, GaSb, InP, InAs and In Sb from 1.5 to 6.0 eV", Phys. Rev. B, 27, 985 (1983).

[19] J. Tauc, in *Amorphous and Liquid Semiconductors,* edited by J. Tauc, Plenum, New York, 1974, p. 159.

[20] D. Bedeaux, J. Vlieger, *Optical properties of surfaces*, Imperial College Press, London, 2004.

[21] P. S. Epstein, "Reflection of waves in an inhomogeneous absorbing medium", Proc. Nat. Acad. Sci, 16, 627 (1930).

[22] R. Diamant, M. Fernandez-Guasti, "Light propagation in 1D inhomogeneous deterministic media: the effect of discontinuities," J. Opt. A: Pure Appl. Opt. 11, 045712 (2009).